\documentstyle[12pt,epsf]{article}
\def\hybrid{\topmargin -20pt    \oddsidemargin 0pt
        \headheight 0pt \headsep 0pt
        \textwidth 6.5in        % US paper
        \textheight 9in         % US paper
        \marginparwidth .875in
        \parskip 5pt plus 1pt   \jot = 1.5ex}

\hybrid
\newcommand{\cA}{{\cal A}}

\newcommand{\cL}{{\cal L}}

\newcommand{\cO}{{\cal O}}
\newcommand{\cP}{{\cal P}}

\newcommand{\cV}{{\cal V}}

\newcommand{\hf}{\frac12}

\newcommand{\bea}{\begin{eqnarray}}
\newcommand{\eea}{\end{eqnarray}}
\newcommand{\be}{\begin{equation}}
\newcommand{\ee}{\end{equation}}
\newcommand{\bt}{\begin{tabular}}
\newcommand{\et}{\end{tabular}}
\newcommand{\ba}{\begin{array}}
\newcommand{\ea}{\end{array}}

\newcommand{\vev}[1]{\langle #1 \rangle}
\newcommand{\Tr}{{\rm Tr}}
\newcommand{\Pf}{{\rm Pf}}
\newcommand{\R}{{\rm Re}}

\newcommand{\Mpl}{M_{\rm Pl}}
\def \jb{{\bar\jmath}}

\def \half{{\textstyle\hf}}

\def \soll={\stackrel{!}{=}}
\def \rf=#1{\stackrel{(\ref{#1})}{=}}

\def \bfm#1{\mbox{\boldmath$#1$}}

\def \Nucl#1{{\em Nucl.~Phys.}\ {\bf B#1}}
\def \NuclProc#1{{\em Nucl.~Phys.~Proc.~Suppl.}\ {\bf #1A}}
\def \PhysR#1{{\em Phys.~Rev.}\ {\bf D#1}}

\def \PhysRep#1{{\em Phys.~Rep.}\ {\bf #1}}
\def \PhysL#1{{\em Phys.~Lett.}\ {\bf #1B}}

\def \Prog#1{{\em Prog.~Theor.~Phys.}\ {\bf #1}}
\def \hep#1{{\tt hep-th/#1}}

\def \nonpert{non\discretionary{-}{}{-}per\-tur\-bative\ }
\def \nonsusy{non\discretionary{-}{}{-}super\-sym\-metric\ }

\def \S{\hat S}
\def \bbar{\overline}

\renewcommand{\thefootnote}{\fnsymbol{footnote}}

\begin{document}
\begin{titlepage}
\begin{center}

\hfill hep-th/9803143\\

\vskip .6in
{\LARGE \bf Quantum modified moduli spaces and 
field dependent gauge couplings\footnote{Research 
 supported by: the DFG (M.K.);
 the NATO, under grant CRG~931380 (J.L.);
 GIF -- the German--Israeli
Foundation for Scientific Research (J.L.).}}
\vskip .5in

{\bf Matthias Klein$^{a,b}$ and  Jan Louis$^{a}$}
\footnote{email: mklein@hera2.physik.uni-halle.de, 
j.louis@physik.uni-halle.de
}
\\
\vskip 0.8cm
{}$^{a}${\em Martin--Luther--Universit\"at 
Halle--Wittenberg,\\
        FB Physik, D-06099 Halle, Germany}
\vskip 0.5cm
{}$^{b}${\em Sektion Physik, 
Universit\"at M\"unchen,\\
Theresienstr.~37, D-80333 M\"unchen, Germany}

\end{center}

\vskip 1.5cm

\begin{center} {\bf ABSTRACT } \end{center}
In this paper we discuss quantum modified moduli spaces
in supergravity. We examine a model suggested by Izawa and Yanagida
and by Intriligator and Thomas that breaks 
global supersymmetry by a quantum 
deformation of the classical moduli space. 
We determine the minimum of the supergravity
potential when the gauge coupling is taken to depend on 
a dynamical field, typically a modulus of string theory.
We find that the only minimum is at the trivial configuration
of vanishing coupling constant and unbroken supersymmetry.
We also discuss models involving more complicated
superpotentials and find that
the gauge coupling is only
stabilized in a supersymmetric ground state. 

\vfill
March 1998
\end{titlepage}

\renewcommand{\thefootnote}{\arabic{footnote}}
\setcounter{footnote}{0}

\section{Introduction}

Supersymmetry plays an important role in particle physics
as a possible extension of the Standard Model.
Supersymmetric field theories
also appear naturally as
the low energy limit of superstring theories.
However, since the physical world at low energies is not supersymmetric, 
any realistic  model of particle physics
necessarily has to incorporate
a mechanism for supersymmetry
breaking.
This breakdown cannot be accomplished by perturbative
quantum corrections but has to occur either at the 
tree level or non-perturbatively. The latter situation
has been considered more attractive (for a recent review see
\cite{PT} and references therein) since it also bears the  possibility of generating
a hierarchy of scales.

Izawa and Yanagida and independently 
Intriligator and Thomas suggested a 
\nonpert mechanism that breaks
supersymmetry for a class of $N=1$
supersymmetric gauge theories \cite{IT}.
These models have a specific matter content 
(e.g.\ $N$ flavors of quark supermultiplets
in the fundamental
representation of an $SU(N)$ gauge theory)
and in the absence of a tree level superpotential 
a moduli space of vacua which is not lifted by 
non-perturbative 
quantum corrections \cite{Seib_D49,qmm1,qmm2}. 
The moduli space can be parametrized by the vacuum expectation values 
(VEVs) of gauge invariant operators 
% effective degrees of freedom 
which satisfy a constraint equation. 
Non-perturbative quantum corrections do not 
lift the moduli space but they do modify
the constraint equation.
As a consequence certain regions (or points)
of the classical moduli space are removed from the
quantum moduli space.
This, in turn, can lead to spontaneous 
supersymmetry breaking whenever the quantum
constraint of the moduli is incompatible with
the classical vacua.

This class of gauge theories can potentially arise 
in the low energy limit of string 
theory \cite{kss,ks,mkL}.
In this case 
the gauge coupling $g$
is not merely a parameter but 
determined by the VEV of a scalar field 
(a string modulus) $\S$.
%If the gauge group is part of the perturbative string spectrum
%this modulus coincides with the string dilaton $S$.
%For non-perturbative factors of the gauge group 
%the corresponding gauge coupling is governed
%by some other modulus; in order to include both
%situations in our discussion  
In string perturbation theory $\S$ is a flat
direction of the effective potential
and hence its VEV $\langle\S\rangle$ is undetermined.
However, \nonpert effects
generically generate a potential for $\S$
and thus do determine $\langle\S\rangle$.
Indeed, the model of \cite{IT} 
contains a potential for $\S$ once the gauge coupling is taken to be field 
dependent. 
However, as we will show in this paper
the only minimum of the potential is 
found for 
vanishing coupling constant $g=0$. 
(This has also been noticed in \cite{MBMQ}.) 
Furthermore, at this point the quantum 
constraint coincides with the classical constraint
and therefore 
supersymmetry is restored at the minimum.
This situation also occurs for a large class
of generalized superpotentials and
thus the mechanism for supersymmetry breaking
suggested in \cite{IT}
appears to be problematic when embedded in string 
theory.

This paper is organized as follows. We first review the mechanism  of \cite{IT} and
indicate the problem of runaway vacua when 
the gauge coupling is taken to 
depend on a dynamical field.
We argue that a
proper treatment of field dependent gauge couplings
requires the coupling of the gauge theory to supergravity. 
This is done in section~3  
using the Veneziano--Yankielowicz formalism 
which includes
a `glueball' superfield in the action
\cite{VY,AKMRV}. Under reasonable assumptions for
the K\"ahler potential for the glueball field
we are able to solve the equations
of motion in the absence of a tree level
superpotential.
For non-vanishing tree-level terms the 
minimization of the supergravity potential 
is only achieved in two (different) perturbative 
expansions.
For the model of Izawa, Yanagida, Intriligator,
and Thomas (IYIT) we find that 
the minimum occurs for 
vanishing gauge coupling. 

In section~4
we extend the analysis to more general superpotentials 
proposed by Dvali and Kakushadze \cite{DK}
which 
show non-trivial minima for $\S$ and therefore stabilize the gauge
coupling at finite values. However, the vacuum is supersymmetric in these
models and the gauge coupling is not small enough to generate
a phenomenologically interesting hierarchy of scales $\Lambda\ll\Mpl$. 
Thus in the class of models considered
$\S$ either runs away to infinity or it is
stabilized in a supersymmetric vacuum without a hierarchy. 

%%%%%%%%%%%%%%%%%%%%%%%%%%%%%%
\section{Supersymmetry breaking on quantum moduli spaces}
Let us first summarize some well known results on the low energy limit of
supersymmetric quantum chromodynamics (SQCD). 
Consider SQCD with gauge group $SU(N_c)$, 
and $N_f$ quark flavors, i.e.\ $2N_f$ chiral 
matter fields $Q^{ri}$,  $\tilde Q_{ri}$,
$r=1,\ldots,N_c,\ i=1,\ldots,N_f$ which are
in the fundamental and  antifundamental representation of 
$SU(N_c)$, respectively.
Affleck, Dine, and Seiberg \cite{ADS}
showed that for $N_f<N_c$ the low energy degrees
of freedom are given by the mesons $M_i^j=\tilde Q_{ri} Q^{rj}$,
and determined the effective (holomorphic)
superpotential to be of the form 
\be \label{W_ADS}
    W=(N_f-N_c)\,\left({\Lambda^{3N_c-N_f}\over\det M}\right)^{1\over N_c-N_f},
\ee
where $\Lambda$ is the dynamically generated SQCD-scale.

For $N_f=N_c$ no superpotential is generated by the strong coupling
dynamics \cite{ADS,Seib_D49} and hence
there is a quantum moduli space spanned by the expectation values of the 
light degrees of freedom. These are most conveniently parametrized by the 
mesons $M_i^j$ and the baryons $B=\det Q$ and $\tilde B=\det\tilde Q$, where
$Q$ and $\tilde Q$ are viewed as $(N_f\times N_c)$-matrices in the 
flavor--color--space.
%$B={1\over N_c!}\epsilon_{r_1\ldots r_{N_c}}
%\epsilon_{i_1\ldots\i_{N_f}}Q^{r_1 i_1}\cdots Q^{r_{N_c}i_{N_f}}$
%and $\tilde B={1\over N_c!}\epsilon^{r_1\ldots r_{N_c}}
%\epsilon^{i_1\ldots\i_{N_f}}\tilde Q^{r_1 i_1}\cdots \tilde
%Q^{r_{N_c}i_{N_f}}$.
However not all of these $N_f^2+2$ degrees of freedom ($M^i_j,B,\tilde B$)
are independent. Classically the fields satisfy $\det M=\tilde B B$;
quantum mechanically this is modified \cite{Seib_D49} to 
\be \label{qconstr} \det M-\tilde B B=\Lambda^{2N_c}.
\ee
Introducing an auxiliary superfield $\cA$ the constraint
(\ref{qconstr}) can be implemented in the superpotential via the Lagrange
multiplier method,
\be \label{W_Seib} W=\cA (\det M-\tilde B B-\Lambda^{2N_c}).
\ee
By adding a large mass term for 
one of the quark flavors 
the superpotential (\ref{W_ADS}) of the corresponding
$N_f=N_c-1$ gauge theory is recovered 
upon integrating out the massive modes.
Furthermore, it can be checked that at any point of moduli space 
the 't Hooft matching conditions hold
for all  unbroken symmetries \cite{Seib_D49}.
An important consequence of the constraint (\ref{qconstr}) is 
the fact that
the chiral symmetry is necessarily broken by the vacuum since 
the expectation
values of the mesons and baryons cannot simultaneously vanish.

Later in this paper
we will concentrate on the case $N_c=2$ since it
is technically easier in
some respects. The fundamental and the antifundamental representation are
equivalent in this case and therefore the $N_f=2$ 
gauge theory  consists of four
fundamental quarks $Q_i, i=1,\ldots,4$. The confined degrees of freedom are
arranged in an antisymmetric Matrix 
$M_{ij}=Q^r_i\epsilon_{rs} Q^s_j$ and there are no baryons.
The constraint corresponding to (\ref{qconstr}) now reads
\be \label{qconstr2} 
\Pf M\equiv{1\over8}\,\epsilon^{ijkl}M_{ij}M_{kl}
=\Lambda^4.
\ee

The authors of \cite{IT} realized that this quantum constraint gives rise to 
a simple supersymmetry breaking mechanism. Consider the $SU(2)\ N_f=2$ model
with six additional singlet fields $X^{ij},i,j=1,\ldots,4$, transforming in 
the conjugate antisymmetric tensor representation
%, i.e.\ $\asb$, 
of the $SU(4)$ flavor group. It is easy to see
that supersymmetry is broken when a term $\rho M_{ij}X^{ij}$ is added to the 
superpotential
%\footnote{It was noticed in \cite{IT} that this 
%mechanism resembles the one proposed by O'Raifeartaigh \cite{O_R}, 
%which has the superpotential
%$W=m\Phi_1\Phi_2+\Phi_0(g\Phi_1\Phi_1-\lambda)$
%and a vacuum with
%spontaneously broken supersymmetry.} 
\be \label{W_IT} W=\rho\,\Tr(MX)+\cA (\Pf M-\Lambda^4),
\ee
where $\rho$ is a 
dimensionless coupling parameter.
The value of the potential $\cV$ of the scalar fields at the (non-supersymmetric)
minimum is \cite{IT} 
\be \label{V_min} \vev\cV=\vev{\sum_i |\partial_i W|^2} = 2|\rho\Lambda^2|^2,
\ee
where $\partial_i$ denotes the derivative with respect to the $i$-th
field.
Note that supersymmetry is restored in the limit $\Lambda\to0$ \cite{MBMQ} which 
corresponds to the trivial situation of an infrared free 
gauge theory. As we will see this generically occurs 
for field dependent gauge 
coupling. In string theory, for example, the 
coupling constant $g$ of the perturbative gauge symmetry is determined by the 
expectation value of the dilaton superfield $S$. More generally, it is known
\cite{smallinst} that at special points of the string moduli space the
gauge symmetry is enhanced by \nonpert effects
 and the coupling of this
\nonpert gauge theory is fixed by a modulus $T$
other than the dilaton $S$. To treat the two
situations simultaneously we denote the modulus
determining the gauge coupling by $\S$. 
At the Planck scale ($\Mpl$) the inverse gauge coupling is matched
to this modulus according to 
$g^{-2}=\R(\S)$ which implies
\be
\label{g2_of_S}
  \Lambda=\Mpl\ e^{-{8\pi^2\S\over b}},
\ee
where $b=3N_c-N_f$.
If this additional superfield $\S$ is present the value of $\Lambda$ in the
vacuum configuration is such that the potential is minimal also with respect to
variations of $\S$. Naively, one could just minimize $\vev\cV$ of 
(\ref{V_min}) to find that the only minimum is at $\S\to\infty$. Thus also in 
the IYIT mechanism we  
encounter the `dilaton problem', that is the 
generic fact that there is no stable vacuum at 
finite dilaton expectation values \cite{DS}.

For a more detailed treatment of this problem we need
some information on the K\"ahler potential $K$ of the theory. 
As the kinetic term of the dilaton is typically of order $\cO(\Mpl^2)$, 
it is necessary to consider the coupling of this gauge theory to supergravity.
The relevant potential of the scalar fields is then given by
\be \label{V_SUGRA}
    \cV=e^{K\Mpl^{-2}}(D_iWg^{i\jb}\,\overline{D_jW}-3\Mpl^{-2}|W|^2),
\ee
 where $D_iW\equiv\partial_iW
+\Mpl^{-2}\,(\partial_iK)\,W$
and $g^{i\jb}$ is the inverse K\"ahler metric, i.e.\   
$g^{i\jb}\,\partial_k\partial_\jb K=\delta^i_k$.

Another question concerns the role of the auxiliary field $\cA$ in 
(\ref{W_Seib}). The Lagrange multiplier technique is a manifestly 
supersymmetric way to enforce the quantum constraint on the moduli. 
But if supersymmetry is broken the condition (\ref{qconstr}) will 
presumably no longer be satisfied.
It seems difficult to determine the deviations from the supersymmetric
quantum constraint in this approach because the Lagrange multiplier is an 
auxiliary field and therefore not dynamical.
A different approach to quantum modified
moduli spaces can be developed using the 
Veneziano--Yankielowicz formalism \cite{VY}
which provides an appropriate 
framework to couple this class of
gauge theories to supergravity.

\section{Coupling to supergravity and field 
dependent gauge couplings}
\subsection{The glueball superfield in global supersymmetry}
Veneziano and Yankielowicz proposed an effective Lagrangian containing
the glueball field $U\equiv{1\over32\pi^2}W^\alpha W_\alpha$, where $W^\alpha$ 
is the field strength chiral superfield. For the case of
`pure glue' ($N_f=0$) they determined
 the superpotential to be \cite{VY}
\be \label{W_VY}
    W=N_c(U\ln{U\over\Lambda^3}-U).
\ee
The supersymmetric minimum is at $U=\omega_k\Lambda^3$ with 
$\omega_k=e^{2\pi ik/N_c}$ a ${\bfm Z}_{N_c}$ phase factor.\footnote{One
has to use the equation of motion $\partial W/\partial U=0$ mod $2\pi i$
because of the different branches of the logarithm. Equivalently this can
be viewed as reflecting the ${\bfm Z}_{N_c}$ symmetry of the Lagrangian.} 
The non-trivial VEV of $U$  corresponds to 
gluino condensation\footnote{The lowest component $\phi$ of the 
superfield $U$ is the gluino bilinear, 
$\phi\equiv(1/32\pi^2)\lambda^\alpha\lambda_\alpha$.}
and the $N_c$ vacua are in accordance with Witten's index \cite{W_index}. 
It is instructing for what follows to discuss the $N_f=0$ case in more detail. 
Subtleties arise when the classical potential of the
scalar fields is considered. 
The authors of \cite{VY} determined the
leading term of the K\"ahler potential to be 
$K\sim (\bbar U U)^{1\over3}$. More generally one can expand the K\"ahler
potential in powers of $1/\Mpl$ and 
on dimensional grounds one has\footnote{By 
using symmetry arguments it can be shown that
this is an expansion in $\Mpl$ and not
in $\Lambda$ \cite{KL}.}
\be \label{Kahler}
    K=\sum_{n=1}^{\infty} c_n{(\bbar U U)^{n/3}\over \Mpl^{2(n-1)}},
\ee
where the $c_n$ are dimensionless real constants.
In global supersymmetry ($\Mpl\to\infty$) one only 
retains the first term in this expansion and denoting by $\phi$ the
lowest component of the chiral multiplet $U$ yields the kinetic terms
\be \label{L_kin_VY}
    \cL_{\rm kin}={c_1\over9}|\phi|^{-4/3}|\partial_\mu\phi|^2+ {\rm fermionic}.
\ee
The potential for the scalar field $\phi$ is given by
\be \label{scalar_pot}
    \cV ={9\over c_1}|\phi|^{4/3}\left|\ln{\phi^{N_c}\over\Lambda^{3N_c}}
                                \right|^2,
\ee
which
shows the $N_c$ minima $\phi=\omega_k\Lambda^3$ and an additional minimum at 
$\phi=0$. In \cite{VY} the latter solution is excluded because the kinetic 
term diverges at $\phi=0$.\footnote{Alternatively this can be seen by a field 
redefinition $\hat U=U^{1\over3}$, such that the kinetic term has the canonical
form. Now $\partial W/\partial\hat U$ shows the additional zero at $\hat U=0$
but at $\hat U=0$ this field redefinition is not invertible.}

The generalisation of (\ref{W_VY}) to $0<N_f<N_c$ is found to be \cite{VY}
\be \label{W_TVY}
    W=U\left(\ln\left({U^{N_c-N_f}\det M\over\Lambda^{3N_c-N_f}}\right)
            -(N_c-N_f)\right).
\ee
This result can also be obtained directly from (\ref{W_VY}) by `integrating in'
quark matter \cite{int_in}.
The vacuum expectation value of $U$ is determined
from $\partial W/\partial U=0$
and found to be
\be 
\vev U=\left({\Lambda^{3N_c-N_f}\over\det M}\right)^{1\over N_c-N_f}.
\ee
Expanding the superpotential in the fluctuations around the vacuum,
$\tilde U=U-\vev U$, shows that the glueball superfield is massive as can be seen from the 
expansion of the superpotential
$W={N_c-N_f\over\vev U}\tilde U^2-(N_c-N_f)\vev U +\cO(\tilde U^3)$.
This was the main reason for the authors of \cite{ADS} not to include
$U$ in the effective Lagrangian, since the Wilsonian effective action only
contains the light degrees of freedom. 
It is easy to see that upon integrating out the massive glueball field
one recovers the Affleck--Dine--Seiberg superpotential (\ref{W_ADS}).

For $N_f=N_c$ one has to include the baryons into 
the analysis and obtains \cite{AKMRV}
(see also \cite{KL})
\be \label{W_Nf}
    W=U\ln\left({\det M-\tilde B B\over\Lambda^{2N_c}}\right).
\ee
The supersymmetric minimum which satisfies
$\partial_iW=0$ results in the constraint (\ref{qconstr}) and
$\vev U=0$. 
The kinetic terms only depend on $K$ and
hence as in the $N_f=0$ case
they diverge at $U=0$
(cf.\ eq.\ (\ref{L_kin_VY})).
However, in this case,
it is not possible to exclude the solution at $U=0$, else there would be no 
stable vacuum at all.
Furthermore, from (\ref{W_Nf}) it is clear that $U$ is now massless
and thus there should be a consistent low energy description containing the
fields $U,M^i_j,B,\tilde B$ in the case $N_f=N_c$. 

The difficulties with singular kinetic terms could be avoided if in the 
general expansion for the K\"ahler potential (\ref{Kahler}) the first two 
coefficients $c_1$ and $c_2$
vanish for $N_f=N_c$. This has the additional property that 
the kinetic term of the glueball field vanishes in the limit of global 
supersymmetry ($\Mpl\to\infty$).
The field $U$ can therefore be identified with the Lagrange multiplier
$\cA$ of eq.\ (\ref{W_Seib}). More precisely,
expanding the logarithm in (\ref{W_Nf}) in powers of 
$\epsilon={\det M-\tilde B B\over\Lambda^{2N_c}}-1$ and comparing with
(\ref{W_Seib}), one sees that $U=\Lambda^{2N_c}\cA$ up to corrections of 
order $\cO(\epsilon^2)$. Therefore the approach
to quantum modified moduli spaces containing the glueball superfield $U$
seems a natural candidate to elevate the Lagrange multiplier in (\ref{W_Seib})
to a dynamical field \cite{qmm2,MBMQ}.
In the following we will assume $c_1=c_2=0$ in 
(\ref{Kahler}) although we were not able to prove this statement more rigorously.

Before turning to a concrete application of this method to supergravity, we
would like to comment on a suggestion of Kovner and Shifman \cite{KS}.
They propose that the $U=0$ solution of the $N_f=0$ theory could be physical, 
in contrast to the discussion of eq.\ (\ref{scalar_pot}) above. 
They argued that the  divergence of the kinetic term simply reflects the fact that one cannot 
trust the Veneziano--Yankielowicz Lagrangian for dynamical issues.
A possible vacuum of 
$SU(N_c)$ gluodynamics with  $\vev U=0$ corresponds to
a vanishing gluino condensate, 
$\vev{\lambda\lambda}=0$.
This has drastic consequences for the vacuum structure of SQCD $(N_f>0)$.
The authors of \cite{ADS} found that the effective superpotential
(\ref{W_ADS}) is generated by gluino condensation (if $N_f<N_c-1$).
The gluino condensate $\vev{\lambda\lambda}=\Lambda^3$ appears in the
strongly coupled unbroken $SU(N_c-N_f)$ pure gauge theory on the Higgs
branch of $SU(N_c)$ SQCD, where all mesons acquire large vacuum 
expectation values. If the chirally symmetric phase exists at $N_f=0$,
then at $N_f>0$ an additional branch with vanishing effective potential
must be present. Consequently, in this case, the quantum moduli space
at $N_f=N_c$ should contain a point at the origin $M=B=\tilde B=0$
in addition to values of $M,B,\tilde B$ obeying (\ref{qconstr}).
Otherwise one would not recover a vanishing superpotential upon
integrating out the massive modes in the limit that all quark flavors 
get very heavy.
This scenario predicts a stable ground state of SQCD for all $N_f\le N_c+1$,
with the vacuum expectation values of all gauge invariant composites
vanishing, in contrast to what was expected from the work of \cite{ADS}. 
On the assumed additional branch of the moduli space the full global symmetry 
$SU(N_f)_L\times SU(N_f)_R\times U(1)_B\times U(1)_R$ is unbroken.
As a consequence the IYIT model presented in the
previous section would not break supersymmetry
but have a supersymmetric ground state at
the origin of the moduli space.

The possible existence of this chirally symmetric phase was also discussed
by the authors of \cite{CM,SZ}. They pointed out that in the $N_f=0$ case
a vanishing gluino condensate does not seem to be compatible with the global
symmetries. A similar problem is found when $N_f>0$; there is no obvious 
spectrum of effective degrees of freedom such that the 't Hooft anomalies
are matched between the microscopic and the macroscopic theory. The effective
(macroscopic) degrees of freedom are gauge invariant holomorphic polynomials
in the elementary fields $Q,\tilde Q,W^\alpha$. For the minimal spectrum
containing only the fields $M_i^j,B,\tilde B,U$ the matching conditions are
not satisfied at the origin of the moduli space (e.g.\ the $SU(N_f)_L^2U(1)_B$ 
anomaly is proportional to $N_c$ and non-zero
in the microscopic theory but vanishes in the macroscopic theory). 
The only possible additional effective degrees of 
freedom are exotics of the form
\bea
  E^{(k)}_B &\sim &\epsilon_{r_1\cdots r_{N_c}}Q^{r_1i_1}\cdots 
                                             Q^{r_{N_c-k}i_{N_c-k}}
       (W^{\alpha_1})^{r_{N_c-k+1}}_{s_1}Q^{s_1i_{N_c-k+1}}\cdots
       (W^{\alpha_k})^{r_{N_c}}_{s_k}Q^{s_ki_{N_c}}\nonumber\\
  E^{(k)}_{M,1} &\sim &\epsilon_{r_1\cdots r_{N_c}}\epsilon^{s_1\cdots s_{N_c}}
            Q^{r_1i_1}\cdots Q^{r_{N_c-k}i_{N_c-k}}
            \tilde Q_{s_1j_1}\cdots\tilde Q_{s_{N_c-k}j_{N_c-k}}
        (W^{\alpha_1})^{r_{N_c-k+1}}_{s_{N_c-k+1}}\cdots 
        (W^{\alpha_k})^{r_{N_c}}_{s_{N_c}}\nonumber\\
  E^{(k)}_{M,2} &\sim &Q^{ri}(W^{\alpha_1}\cdots W^{\alpha_k})^s_r\tilde Q_{sj}
        .\nonumber
\eea  
For $N_f<N_c$ 
it seems difficult to find a subset of these that satisfy all of the matching
conditions. For $N_f=N_c$ the argument can be made more precise. Since the
non-anomalous R-charge of the quarks vanishes in this case, the R-charge of
the fermionic component of $E^{(k)}_{B/M}$ is $\ge 1$. However the $U(1)_R$ 
anomaly is
matched between the microscopic and the macroscopic theory when only the
the effective fields $M_i^j,B,\tilde B,U$ are present. This matching is no
longer possible when any of the exotics is added, because there are no
(holomorphic) exotics with zero or negative R-charge.
We will therefore assume that at least for $N_f=N_c$ there is no chirally 
symmetric phase.

\subsection{Quantum moduli spaces in supergravity}
As an application of the ideas presented above let us now discuss the
embedding of quantum modified moduli spaces in supergravity. 
We start with the situation where 
no tree-level superpotential is present. For an $SU(2)$
gauge theory with $N_f=2$ quark flavors the superpotential 
is given by
\be W=U\ln{\Pf M\over\Lambda^4}.
\ee

As we do not know the K\"ahler potential of the theory for $M$ and $U$, we 
make an Ansatz as a power series in the fields:
\be \label{Kahler_qmm}
 K=\sum_{i<j}\sum_{n=1}^{\infty} a_n
     {(\bbar M_{ij}M_{ij})^{n/2}\over \Mpl^{2(n-1)}}
   \ +\ \sum_{n=3}^{\infty} c_n{(\bbar U U)^{n/3}\over \Mpl^{2(n-1)}}
   \ -\ d\ln(\S+\bar{\S}),
\ee
where the coefficients $a_n,c_n,d$ are real dimensionless constants. Positivity
of the K\"ahler metric $\partial_i\partial_\jb K$ requires that generically
at least $a_1,c_3,d$ are $>0$. 
If $\S$ is the dilaton, then $d=1$ holds.

The supersymmetric vacuum is determined by the 
equations  
\bea \label{susyeq}
     D_{M_{ij}}W &= &{U\over\Pf M}\half\epsilon^{ijkl}M_{kl}
        +{K_{M_{ij}}\over\Mpl^2}U\ln{\Pf M\over\Lambda^4}\ \soll=0,\nonumber\\
     D_UW &= &\ln{\Pf M\over\Lambda^4}+{K_U\over\Mpl^2}
     U\ln{\Pf M\over\Lambda^4}\ \soll=0,\\
     D_{\S}W &= &8\pi^2\,U-{d\over2\R(\S)}U\ln{\Pf M\over\Lambda^4}
     \ \soll=0,\nonumber
\eea
where $K_{M_{ij}}\equiv\partial K/\partial M_{ij}$, 
$K_U\equiv\partial K/\partial U$, and we have used the relation 
$\Lambda^4=\Mpl^4\,e^{-8\pi^2\S}$. 
The eqs.\ (\ref{susyeq}) are satisfied for
$\Pf M=\Lambda^4,\ U=0$ and $\S$ arbitrary
which determines the supersymmetric vacuum configurations.
The supergravity potential $\cV$, defined in (\ref{V_SUGRA}) vanishes
at this configuration, since $D_iW=W=0$. To see if there are other minima
than the supersymmetric one, we need the explicit expression of $\cV$.
It is given by \footnote{For simplicity of notation we use the same symbols 
for chiral superfields and their lowest components.}
\bea \cV &= &e^{K\Mpl^{-2}}\left(\left|\ln{\Pf M\over\Lambda^4}\right|^2
                    \tilde g^U\Mpl^4\ +\ |U|^2\left(\sum_{i<j}\left|
                    {\half\epsilon^{ijkl}M_{kl}\over\Pf M}
                    +{K_{M_{ij}}\over\Mpl^2}\ln{\Pf M\over\Lambda^4}\right|^2
                    g^{M_{ij}\bar M_{ij}}\right.\right.\nonumber\\
     && \hspace*{-1.3cm}
         +{d\over\Mpl^2}\left|{16\pi^2\over d}\R(\S)-\ln{\Pf M\over\Lambda^4}
                               \right|^2
         \left.\left.+{1\over\Mpl^2}\left({|U|^2\over\Mpl^6}(\tilde K_U)^2
                                     \tilde g^U+2\tilde K_U\tilde g^U-3\right)
            \left|\ln{\Pf M\over\Lambda^4}\right|^2\right)\right),
\eea
where $\tilde K_U=\sum_{n=3}^\infty c_n{n\over3}\left({|U|^{2/3}\over\Mpl^2}
\right)^{n-3}$ and $\tilde g^U=\Mpl^{-4}g^{U\bar U}=\sum_{n=1}^\infty
\gamma_n\left({|U|^{2/3}\over\Mpl^2}\right)^{n-1}$ are dimensionless real 
functions of $U$, and $\tilde g^U>0$. 
An exact minimization of this potential is
difficult but one can show that $\cV>0$ holds if $\Pf M\neq\Lambda^4$
when reasonable assumptions on the K\"ahler potential are made.
 We split the potential into two parts, one containing all
the terms proportional to $|\ln(\Pf M/\Lambda^4)|^2$ and another one 
consisting of the remaining terms. For the first part one has
\be \label{V_pos}
 {1\over\tilde K_U}\left({(|U|^2\tilde K_U)^2\over\Mpl^8}\tilde K_U\tilde g^U
 +{|U|^2\tilde K_U\over\Mpl^2}(2\tilde K_U\tilde g^U-3)
 +\tilde K_U\tilde g^U\Mpl^4\right)>0
 \qquad{\rm if\ }\tilde K_U\tilde g^U>{3\over4}.
\ee
The second part is always positive and therefore $\cV$ is positive whenever
$\tilde K_U\tilde g^U>{3\over4}$. We do not know the coefficients $c_n$ in
the general expansion of $\tilde K_U$ but when we assume that the coefficients
$\tilde c_n$ of the K\"ahler metric 
$\partial_U\partial_{\bar U}K=\sum_n\tilde c_n(|U|^{2/3}/\Mpl^2)^n$
are at most of order one, i.e.\ do not grow with $n$, (else it would diverge 
for $U\approx\Mpl^3$), then $c_n=(3/n)^2\tilde c_n=\cO(9/n^2)$. From the power
series expansion of $\tilde K_U$ and $\tilde g^U$ one finds that the
condition $\tilde K_U\tilde g^U>{3\over4}$ holds if $|U|<0.5\,\Mpl^3$.
In addition, a numerical analysis shows that the left hand side of
(\ref{V_pos}) is always $>-|U|^2$ (when the K\"ahler metric has no 
singularities for $0<|U|<1$). In addition, one has for the second 
part of the potential
$$\sum_{i<j}\left|{\half\epsilon^{ijkl}M_{kl}\over\Pf M}
  +{K_{M_{ij}}\over\Mpl^2}\ln{\Pf M\over\Lambda^4}\right|^2
  g^{M_{ij}\bar M_{ij}}+{d\over\Mpl^2}\left|{16\pi^2\over d}\R(\S)
  -\ln{\Pf M\over\Lambda^4}\right|^2\ >\ 
  \left|\ln{\Pf M\over\Lambda^4}\right|^2$$
and therefore $\cV>0$.
This implies that the supersymmetric configuration is indeed the absolute
minimum of $\cV$.
Thus, as could be expected, the quantum moduli space of global
supersymmetry is not altered by supergravity.

\subsection{Local supersymmetry breaking with fixed $\Lambda$}
Now we are ready to couple the IYIT model 
%described above 
to supergravity. Before discussing the question of a field dependent
scale $\Lambda$ let us determine the effects of supergravity when $\Lambda$ 
is a fixed parameter of the theory. We first
calculate the correction for the constraint (\ref{qconstr2}) 
when supersymmetry is broken. In the limit 
$\Mpl\to\infty$ this constraint is not modified because $U$ is an auxiliary 
Lagrange multiplier field in this case. The corrections to the ratio 
$\Pf M/\Lambda^4$ will therefore be suppressed by some power of
$|\Lambda|/\Mpl$. We assume that the expectation values of all fields are 
of order $\cO(\Lambda)$ and  $|\Lambda|\ll\Mpl$. 
The superpotential (\ref{W_IT}) now reads
\be \label{W_ITVY}
    W=\rho\,\Tr(MX)+U\ln{\Pf M\over\Lambda^4}.
\ee
At the leading order in $|\Lambda|/\Mpl$ the K\"ahler potential is
\be
  K=\sum_{i<j}a_1(\bbar M_{ij}M_{ij})^{1/2}\ +\ 
    \sum_{i<j}b_1(\bbar X_{ij} X_{ij})
     \ +\ \cO\left({|\Lambda^2|\over\Mpl^2}\right).
\ee
Next, the potential (\ref{V_SUGRA}) is expanded in powers of $|\Lambda|/\Mpl$:
\bea \label{pot}
 \cV &= &\alpha_1\sum_{i<j}\left|\rho X^{ij}+{U\over\Pf M}\half\epsilon^{ijkl}
            M_{kl}\right|^2|M_{ij}|\ +\ \beta_1|\rho|^2\sum_{i<j}|M_{ij}|^2\\
     &&\hspace{-1cm}+\left|\ln{\Pf M\over\Lambda^4}\right|^2
            \left(\gamma_1(\Mpl^4+K\Mpl^2+\half K^2)
       \ +\ \gamma_2(\Mpl^2+K)|U|^{2/3}+\gamma_3|U|^{4/3}\right)
       \ +\ \cO\left({|\Lambda^6|\over\Mpl^2}\right),\nonumber
\eea
where $\alpha_1={4\over a_1}$, $\beta_1={1\over b_1}$, $\gamma_1={1\over c_3}$,
$\gamma_2=-{16\,c_4\over9\,c_3^2}$, $\gamma_3={256\,c_4^2\over81\,c_3^3}
-{25\,c_5\over9\,c_3^2}$.
The minimum in the directions $X^{ij}$ and $U$ is at $X^{ij}=U=0$.
Then, minimizing with respect to $M_{12}$ yields
\bea 
\beta_1|\rho|^2 M_{12}\ +\ {a_1\over2}\gamma_1{M_{12}\over|M_{12}|}(\Mpl^2+K)
         \left|\ln{\Pf M\over\Lambda^4}\right|^2\nonumber\\
     \ +\ \gamma_1{\bbar M_{34}\over\Pf\bbar M}(\Mpl^4+K\,\Mpl^2+\half K^2)
          \ln{\Pf M\over\Lambda^4}
     \ +\ \cO\left({|\Lambda^4|\over\Mpl^2}\right) &= &0.
\eea
This quadratic equation in $\ln(\Pf M/\Lambda^4)$ is solved by
\bea   \ln{\Pf M\over\Lambda^4} &= &-|\rho|^2{\beta_1\over\gamma_1}
       {M_{12}\over\bbar M_{34}}{\Pf\bbar M\over\Mpl^4}\ +\ 
       \cO\left({|\Lambda^6|\over\Mpl^6}\right) \nonumber\\
    &= &\ -|\rho|^2{\beta_1\over\gamma_1}\hf\sum_{i<j}{|M_{ij}|^2\over\Mpl^4}
    \ +\ \cO\left({|\Lambda^6|\over\Mpl^6}\right).
\eea
The second equality follows from the other equations 
$\partial\cV/\partial M_{ij}=0$; they imply $|M_{12}|=|M_{34}|$,
$|M_{13}|=|M_{24}|$, $|M_{14}|=|M_{23}|$ and the phases are such that
$e^{-i\varphi}\Pf M=\half\sum_{i<j}|M_{ij}|^2$, where $\varphi$ is defined by
$M_{12}M_{34}=e^{i\varphi}|M_{12}M_{34}|$.
Therefore the leading  correction to the 
constraint (\ref{qconstr2}) is given by
\be \label{correc}
     {\Pf M\over\Lambda^4}=1\ -\ |\rho|^2e^{-i\varphi}{\Lambda^4\over\Mpl^4}
       \ +\ \cO\left({|\Lambda^6|\over\Mpl^6}\right),
\ee
where we have assumed 
$\beta_1=\gamma_1=1$ for simplicity. 

At the first order in $|\Lambda^2|/\Mpl^2$ the result obtained in global 
supersymmetry for the vacuum energy $\vev\cV$, eq.\ (\ref{V_min}), 
is only modified by the prefactor $e^{K\Mpl^{-2}}$,
\be \label{V_min2} 
    \vev\cV=2|\rho\Lambda^2|^2\,\left(1+2a_1{|\Lambda^2|\over\Mpl^2}\right)
     \ +\ \cO\left({|\Lambda^4|^2\over\Mpl^4}\right).
\ee
The correction (\ref{correc}) causes a shift of order 
$\cO({|\Lambda^4|^2\over\Mpl^4})$ in $\cV$ but there are other corrections
of the same order due to the fact that the vacuum configuration is no longer 
at $X^{ij}=U=0$ when higher powers of $|\Lambda|/\Mpl$ are taken into account 
in (\ref{pot}). To determine these would take more effort. 
Instead let us now turn to the case of a
field-dependent scale $\Lambda$.

%The value of $\cV$ at the minimum is
%\be \cV=|m|^2|\Pf M|\left(2+|m|^2{|\Pf M|\over\Mpl^4}\right)
%     =2|m|^2|\Lambda^4|\left(1-|m|^2{\R(\Lambda^4)-\hf|\Lambda^4|\over\Mpl^4}
%        \right).
%\ee

\vskip 1.5cm
\subsection{Field dependent gauge coupling}
In the previous section we discussed quantum modified moduli in supergravity
when either the tree-level superpotential vanishes or the scale $\Lambda$ is
a fixed parameter. Let us now treat in more detail the model of \cite{IT} with
$\Lambda$ being a field dependent scale.
In particular we are interested if there is a stable 
vacuum at finite values of $\S$ or not.

We need to locate the minimum of the potential $\cV$ associated to the
superpotential (\ref{W_ITVY}). 
Since the analytical expression of $\cV$ is again
rather involved it is difficult to minimize the supergravity 
potential exactly. When discussing the case of fixed $\Lambda$ we 
expanded the potential in powers of $|\Lambda|/\Mpl$.
The same expansion procedure now 
produces at leading
order
\be\label{neweq}
    \vev\cV=|\rho\Lambda^2|^2\,\R(\S)^{-d}
     \,\left(1+2a_1{|\Lambda^2|\over\Mpl^2}\right)
     \ +\ \cO\left({|\Lambda^4|^2\over\Mpl^4}\right),
\ee 
where $\Lambda^4=\Mpl^4\,e^{-8\pi^2\S}$.
The additional term $g^{\S\bar{\S}}|D_{\S} W|^2$ in the potential $\cV$ 
is of order $\cO(|\Lambda^4|^2/\Mpl^4)$ because $X^{ij}$ and $U$ vanish 
at leading order and the additional term $-d\ln(\S+\bar{\S})$
in the K\"ahler potential gives the  overall
factor $1/2\R(\S)^d$ in (\ref{neweq}).
Positivity of the K\"ahler metric in the limit $\Mpl\to\infty$
requires $a_1>0$. Thus we recover the problem of the runaway vacuum 
indicated above: the only  minimum is at $\S\to\infty$
which is again a supersymmetric solution.

It is interesting to note that, at the leading order, this result
is not altered when an arbitrary superpotential
$W_N(\S)$ is added to (\ref{W_ITVY}), which only depends on $\S$ and is at most of order 
$\cO(\Lambda^3)$. Since 
$g^{\S\bar{\S}}|D_{\S}(W_{\rm IYIT}+W_N)|^2$
and $\Mpl^{-2}|W_{\rm IYIT}+W_N|^2$
are of order
$\cO(|\Lambda^6|/\Mpl^2)$ we obtain
$\vev\cV=|\rho\Lambda^2|^2\,\R(\S)^{-d}+\cO(|\Lambda^6|/\Mpl^2)$.
This implies that 
the string modulus $\S$ is still driven to infinity
even in the presence of
an arbitrary superpotential 
$W_N(\S)$.

An expansion in $|\Lambda|/\Mpl$ is  
slightly more problematic when the gauge coupling 
depends on the dynamical field $\S$, since the ratio
$\Lambda/\Mpl=e^{-2\pi^2\vev\S}$ is fixed by the equations of motion and
not apriori a small parameter.
Thus within a self-consistent treatment we can only state 
that for large (but finite) $\vev\S$ there is no stable vacuum.
The domain which is not covered by this approximation corresponds
to a gauge theory which is strongly coupled at $\Mpl$.
For example,  demanding 
$4\pi/g^2>1$ results in $|\Lambda^2|/\Mpl^2 < 0.04$ which does justify
the above expansion procedure.

%Including  the next order in eq.\ (\ref{V_min2})
%does not change this conclusion since
%the expansion in powers of $|\Lambda^2|/\Mpl^2$ only makes sense, if 
%it turns out that there is a  minimum for $\S$ at values 
%$0<e^{-4\pi^2\R(\S)}\ll 1$, i.e.\ $1\ll 4\pi^2\R(\S)<\infty$.
%As we did not find
%such a minimum we cannot aposteriori justify the expansion in powers
%of $|\Lambda^2|/\Mpl^2$, when $\Lambda$ is considered to depend on $\S$.

An alternative approximation which does not use
the smallness of $|\Lambda^2|/\Mpl^2$
 is an expansion of $\cV$ in powers of $\rho$.
%At least for $\rho\ll 1$ this should give us a good approximation to the
%solution of the problem. 
In section 3.2 we found the exact solution for 
the $\rho=0$ case. Let us now  
%and for $0\neq\rho\ll 1$ we 
calculate the perturbations around the $\rho=0$
solution. For simplicity we place ourselves at the point $M^{(0)}_{ij}$ 
of the moduli space where $M^{(0)}_{12}=M^{(0)}_{34}=\Lambda^2,\ 
M^{(0)}_{13}=M^{(0)}_{14}=M^{(0)}_{23}=M^{(0)}_{24}=0$.
The Ansatz
\be \label{ansatz}
M_{ij}=M^{(0)}_{ij}+\rho\tilde M_{ij},\qquad U=\rho\tilde U,
\ee
\medskip
yields the superpotential
\be W=\rho\Lambda^2(X^{12}+X^{34})+\cO(\rho^2).
\ee
\medskip
To obtain an explicit expression for the scalar potential we need an 
additional assumption on the form of the K\"ahler potential. In analogy to
(\ref{Kahler_qmm}) we make the Ansatz
\medskip
\be \label{Kahler_IT}
 K=\sum_{i<j}\sum_{n=1}^{\infty} a_n
     {(\bbar M_{ij}M_{ij})^{n/2}\over \Mpl^{2(n-1)}}
   \ +\ \sum_{i<j}\sum_{n=1}^{\infty} b_n
     {(\bbar X^{ij}X^{ij})^n\over \Mpl^{2(n-1)}}
   \ +\ \sum_{n=3}^{\infty} c_n{(\bbar U U)^{n/3}\over \Mpl^{2(n-1)}}
   \ -\ d\ln(\S+\bar{\S}).
\ee
\bigskip

For $\cV$ of (\ref{V_SUGRA}) this gives to lowest order in $\rho$
\bea \label{V_IT}
\nobreak
    \cV &= &e^{K\Mpl^{-2}}|\rho|^2\left(\left(
      \left|X^{12}+{\tilde U\over\Lambda^2}+(X^{12}+X^{34})\tilde K_M\right|^2
      \right.\right.\nonumber\\
    &&\left.+\left|X^{34}+{\tilde U\over\Lambda^2}+(X^{12}+X^{34})\tilde K_M
       \right|^2\right)\tilde g^M\Mpl^2\ +\ \left|{\tilde M_{12}+\tilde M_{34}
        \over\Lambda^2}\right|^2g^{U\bar U}\\
    &&+|\Lambda^2|^2\left(\left|1+(X^{12}+X^{34}){K_{X^{12}}\over\Mpl^2}
                            \right|^2g^{X^{12}\bar X^{12}}
       +\left|1+(X^{12}+X^{34}){K_{X^{34}}\over\Mpl^2}
                            \right|^2g^{X^{34}\bar X^{34}}\right)\nonumber\\
    &&+{d\over\Mpl^2}\left|{16\pi^2\over d}\tilde U\R(\S)-\Lambda^2
       (X^{12}+X^{34})\right|^2-3{|\Lambda^2|^2\over\Mpl^2}|X^{12}+X^{34}|^2
       \Bigg)\ +\ \cO\left(|\rho|^4\right),\nonumber
\eea
where $\tilde K_M=\sum_{n=1}^\infty a_n{\textstyle{n\over2}}
       \left({|\Lambda^2|\over\Mpl^2}\right)^n$ and 
      $\tilde g^M=\Mpl^{-2}g^{M_{12}\bar M_{12}}|_{M_{12}=\Lambda^2}
        =\sum_{n=1}^\infty\alpha_n\left({|\Lambda^2|\over\Mpl^2}\right)^n$.

The minimum with respect to $M_{ij}$ and $U$ is at
\bea \label{h_of_S}
     &&\tilde M_{12}=-\tilde M_{34},\qquad 
       \tilde U= -(X^{12}+X^{34})\Lambda^2\,h(\S)\\
     {\rm with}&&h(\S)={\tilde g^M(\half+\tilde K_M)
                        -8\pi^2\R(\S){|\Lambda^2|^2\over\Mpl^4}\over
             \tilde g^M+{2\over d}(8\pi^2\R(\S))^2{|\Lambda^2|^2\over\Mpl^4}}.
    \nonumber
\eea

To further simplify the expression of $\cV$ we will assume canonical 
K\"ahler potential for the $X^{ij}$, i.e\ $K_{X^{12}}=\bar X^{12}$, 
$K_{X^{34}}=\bar X^{34}$. This is justified because the $X^{ij}$ are
elementary fields and not composite and the one-loop correction to the 
canonical K\"ahler potential is of order $\cO(|\rho|^2)$ \cite{BHOO}. 
Now, inserting (\ref{h_of_S}) in (\ref{V_IT}) gives
\bea \cV &= &e^{K\Mpl^{-2}}|\rho|^2\Bigg(\left(
     \left|X^{12}+(X^{12}+X^{34})(\tilde K_M-h(\S))\right|^2\right.\nonumber\\
     &&\left.+\left|X^{34}+(X^{12}+X^{34})(\tilde K_M-h(\S))\right|^2
       \right)\tilde g^M\Mpl^2\nonumber\\
     &&+\left(2+{2\over\Mpl^2}|X^{12}+X^{34}|^2+{1\over\Mpl^4}|X^{12}+X^{34}|^2
        \left(|X^{12}|^2+|X^{34}|^2\right)\right)|\Lambda^2|^2\nonumber\\
     &&+\left(d\left({16\pi^2\over d}h(\S)\R(\S)+1\right)^2-3\right)
                                                         |X^{12}+X^{34}|^2
        {|\Lambda^2|^2\over\Mpl^2}\Bigg)\ +\ \cO\left(|\rho|^4\right).
\eea

If $h(\S)\ge0$ then the minimum is at is at $X^{12}=X^{34}=0$ (we assume 
$d\ge1$). This is indeed the case if the coefficients $a_n$ of the
K\"ahler potential are at most of order $\cO(4/n^2)$, which is necessary for
a regular behavior of the K\"ahler metric 
$\partial_{M_{ij}}\partial_{\bar M_{ij}}K$. This result is obtained 
from a numerical analysis of the function $h(\S)$ defined in (\ref{h_of_S}).

On the subspace of field configurations where all fields except $\S$ are
at the minimum values $\cV$ has the form
\be \cV|_{\rm min}= |\rho|^2\Mpl^4{\exp\left(2\sum_{n=1}^\infty 
                  a_ne^{-4\pi^2\R(\S)n}-8\pi^2\R(\S)\right)
                 \over\R(\S)^d}
    \ +\ \cO(|\rho|^4).
\ee
We find again that, at least to order $\cO(|\rho|^2)$\,\footnote{This should 
be a good approximation because in the IYIT model global supersymmetry is 
broken for arbitrarily small $\rho$.}
and for $\tilde K_M\ge0$, there is no stable ground
state except the trivial one at $\S\to\infty$.
To summarize, the supersymmetry breaking mechanism 
proposed in  \cite{IT} only works when the scale $\Lambda$ is a fixed 
parameter. When this mechanism is embedded
in string theory, the gauge coupling 
and therefore the SQCD-scale depend on a modulus field $\S$ and the vacuum is 
driven to the trivial case $\Lambda=0$.

\section{Other models and conclusion}
An interesting question to ask is whether the behavior of the simple IYIT 
model is generic or if there are more general models that break supersymmetry 
on quantum moduli spaces and at the same time stabilize the string modulus 
$\S$ at finite expectation values. We start by considering more general
tree-level superpotentials and then show how this can be generalized to
other gauge groups.

Let us first mention a simple mechanism to fix the VEV
of $\S$ while 
{\em preserving} supersymmetry which was proposed by Dvali and Kakushadze 
\cite{DK}.
Consider an $SU(2)$ gauge theory with $N_f=2$ and one additional singlet field
$X$ and superpotential
\be  \label{W_DK}
   W = (\xi-X)\, \Pf M+{\eta\over k+1}\,X^{k+1}
+U\ln{\Pf M\over\Lambda^4} .
\ee
It
results in a supersymmetric ground state at
\be \label{glob_stable}
    \Lambda^4=\Pf M=\eta\,\xi^k\ \to\ 
    \S=-(k/8\pi^2)\log(\xi\eta^{1\over k}), 
\ee
provided $|\eta\,\xi^k|<1$, which is equivalent to $1/g^2=\R(\S)>0$.
The solution (\ref{glob_stable}) is obtained by 
demanding $\partial_iW=0$
%the global equations of motion
 \cite{DK}. A numerical analysis
shows that when supergravity is switched on there is still a supersymmetric
solution, i.e.\ $D_iW=0$. For example, for $\eta=1,\xi=0.3,k=2$ one finds%
\footnote{To do this calculation we assumed canonical K\"ahler potential
for $X$ and chose the coefficients for $M_{ij}$ and $U$ in (\ref{Kahler_IT}) 
to be $c_3=1$, $a_1=4$, $a_2=1$, $a_3=4/9$, $a_4=1/4$.
Higher orders do not significantly change the results. Different values
for the coefficients $a_n$, which still decrease like $1/n^2$ to guarantee
the regularity of the K\"ahler metric, may change the results by 10--20\%.} 
$\Lambda^4\approx\Pf M\approx 0.16\,\Mpl^4\ \to\ 8\pi^2\S\approx 1.8$. The 
supergravity corrections to the quantum constraint are very small
$1-\Pf M/\Lambda^4\approx 7\cdot 10^{-6}$, but the result for $\Pf M$ differs
considerably from the one obtained in global supersymmetry which is given by 
$\Pf M=\eta\,\xi^2=0.09\,\Mpl^4$.
As the superpotential $W$ does not vanish at the minimum ($W\approx 0.005\,
\Mpl^3$), the potential $\cV\approx -9\cdot 10^{-4}\,\Mpl^4$ is negative 
for this non-trivial ground state, whereas $\cV=0$ in the limit $\S\to\infty$.

This can easily be generalized to a large class of models achieving the
same results \cite{DK}. To get dimensionless couplings in more general 
superpotentials it is convenient to scale out factors of $\Mpl$ 
and define
$\hat M_{ij}=M_{ij}/\Mpl^2$. Let $\cP(\Pf\hat M)$ be an arbitrary 
polynomial of $\Pf\hat M$ and $X$ a $SU(2)$ singlet. Then consider
\be W=\rho X\cP(\Pf\hat M)\,\Mpl^2+U\ln{\Pf M\over\Lambda^4}.
\ee
As there is no potential for the singlet field $X$, it acts like a Lagrange
multiplier enforcing another constraint on the moduli (in addition to the
quantum constraint). 
The global equations of motion set $X=U=0=\cP(\Pf\hat M)$ and 
$\Pf M=\Lambda^4$. Therefore $\Lambda^4$ is fixed to be one of the zeroes of 
$\cP$. If $\cP(x)$ has zeroes at values $|x|=e^{-8\pi^2\R(\S)}<1$ then $\S$ 
and therefore also the gauge coupling are
stabilized at finite values, or else supersymmetry is
broken. In the former case, because of $\vev W=0$ the globally 
supersymmetric solution, $\partial_iW=0$, is still a solution of $D_iW=0$ in 
supergravity. To see if there are additional \nonsusy minima of
the potential, we have to write down the explicit expression for $\cV$.
We expand in powers of $\rho$ and make the Ansatz (\ref{ansatz}). At lowest
order in $\rho$ this yields the superpotential
\be W=\rho X\cP\left({\Lambda^4\over\Mpl^4}\right)\,\Mpl^2
    \ +\ \cO\left(\rho^2\right)
\ee
and the scalar potential (we assume $K_X=\bar X$)
\bea \cV &= &|\rho|^2e^{K\Mpl^{-2}}\left(2\left|X\left(\cP^\prime\,
             {\Lambda^4\over\Mpl^4}+\cP\,\tilde K_M\right)
            +{\tilde U\over\Mpl^2}\right|^2\Mpl^2\tilde g^M
             {\Mpl^4\over|\Lambda^2|^2}\right.\nonumber\\
        &&+|\cP|^2\left(\Mpl^4+2|X|^2\Mpl^2+|X|^4\right)
           +\left|{\tilde M_{12}+\tilde M_{34}\over\Lambda^2}\right|^2
            g^{U\bar U}
           \nonumber\\
        &&\left.+\Mpl^2 d\left|{\tilde U\over\Mpl^2}{16\pi^2\over d}
                                                    \R(\S)-X\cP\right|^2
          -3\Mpl^2|X\cP|^2\right)\ +\ \cO\left(|\rho|^4\right),
\eea
where $\cP^\prime(x)=d\cP(x)/dx$. 
An analogous discussion as in section 4 shows that the minimum is at
\bea &&\tilde M_{12}=-\tilde M_{34},\qquad 
       \tilde U= -X\cP\,\Mpl^2\,h(\S)\\
     {\rm with}&&h(\S)={\tilde g^M({\cP^\prime\,\Lambda^4\over\cP\,\Mpl^4}
                                   +\tilde K_M)
                        -8\pi^2\R(\S){|\Lambda^2|^2\over\Mpl^4}\over
           \tilde g^M+{2\over d}(8\pi^2\R(\S))^2{|\Lambda^2|^2\over\Mpl^4}}.
    \nonumber
\eea
Generically the function $h(\S)$ is positive and, as a consequence, the minimum
is at $X=0$. Thus we obtain
\be \label{VP}
\cV|_{\rm min}= |\rho|^2\Mpl^4\, |\cP(e^{-8\pi^2\S})|^2\, 
  {\exp\left(2\sum_{n=1}^\infty a_ne^{-4\pi^2\R(\S)n}\right)\over 2\R(\S)^d}\, 
   \ +\ \cO(|\rho|^4).
\ee
This result remains true in the limit $\cP\to0$, although $h(\S)$ is not
well defined in this limit. Thus the supersymmetric configuration
$\cP=0=X=U$ is the absolute minimum of $\cV$
with vanishing vacuum energy.
For large  $\R(\S)$ there are no additional \nonsusy relative minima if $\cP$ 
contains no constant terms (i.e.\ $\cP(0)=0$).
The minimum requires $\tilde K_M+1/(8\pi^2\R(\S)^d)+\cP^\prime/\cP=0$, but 
since for large $\R(\S)$ the quotient $\cP^\prime/\cP$ goes like 
$e^{+8\pi^2\S}$ this condition can never be satisfied.

The following class of models seems to be interesting, because they allow to 
stabilize the gauge coupling without introducing additional singlet fields 
at all. Let $\cP(\Pf\hat M)$ again be an arbitrary polynomial of 
$\Pf\hat M$ and consider the superpotential
\be W=\rho\,\cP(\Pf\hat M)\,\Mpl^3+U\ln{\Pf M\over\Lambda^4}.
\ee
In global supersymmetry and fixed $\Lambda$ this has a supersymmetric
minimum at $\Pf M=\Lambda^4$ and $U=-\rho\,\cP^\prime\,\Lambda^4/\Mpl$.
Varying with respect to $\S$ demands $U=0$ and therefore fixes 
$\Lambda^4/\Mpl^4$ to be a zero of $\cP^\prime$.
If $\cP^\prime(x)$ has no zero at $|x|<1$, then only the trivial solution,
$\S\to\infty$ and $\Pf M=0$, remains.
Of course a correct treatment of this problem requires to solve the 
supergravity equations of motion. Numerically one finds that there is
a supersymmetric solution, i.e.\ $D_iW=0$, and at least for small $\rho$,
$\Lambda^4/\Mpl^4$ lies near a zero of $\cP^\prime$. E.g.\ for 
$\cP(x)=2x^2-x$ and $\rho=0.1$ one has $e^{-8\pi^2\S}\approx0.254$. 
The quantum constraint gets only small corrections, 
$1-\Pf M/\Lambda^4\approx6\cdot 10^{-5}$, 
and the vacuum energy is negative $\vev\cV\approx -0.02\,\Mpl^4$.

The analytic expression of the potential at the minimum with respect to $U$ 
and $M_{ij}$ is given by 
\bea \cV|_{\rm min} &= &|\rho|^2e^{K\Mpl^{-2}}\Mpl^4\,\Bigg(2\left|\cP^\prime\,
                     {\Lambda^4\over\Mpl^4}+\cP\,\tilde K_M-\cP h(\S)\right|^2
                     {\Mpl^4\over|\Lambda^2|^2}\tilde g^M \nonumber\\
         &&+d\left|(16\pi^2/d)\R(\S)\cP h(\S)+\cP\right|^2-3|\cP|^2\Bigg)
           \ +\ \cO\left(|\rho|^4\right).
\eea
In general it is rather hard to see if there are additional minima which
possibly stabilize $\S$ while breaking supersymmetry. However the simplest
case $\cP(\Pf\hat M)=\Pf\hat M$ can be treated at least numerically.
As $\cP^\prime$ has no zeroes we expect that there is no non-trivial
vacuum configuration. The supergravity potential now reads
(for simplicity we assume $d=1$)
\bea
 \cV|_{\rm min} &=& |\rho|^2e^{K\Mpl^{-2}}{|\Lambda^4|^2\over\Mpl^4}\,
   \left(2\left(1+\tilde K_M-h(\S)\right)^2{\Mpl^4\over|\Lambda^2|^2}\tilde g^M
   +|16\pi^2\R(\S)h(\S)+1|^2-3\right)\nonumber\\
&& +\cO\left(|\rho|^4\right).
\eea
We find that $\cV$ is positive and monotonously decreasing.\footnote{This was 
obtained by a numerical analysis taking for $\tilde K_M$, $\tilde g^M$
the power series expansion given in the line below eq. (\ref{V_IT}),
expanding $\cV$ in powers of $|\Lambda^2|/\Mpl^2$ up to 
$(|\Lambda^2|/\Mpl^2)^8$ and assuming for the coefficients of the K\"ahler
potential $a_n=4/n^2$. If the coefficients are smaller, e.g.\ $a_n=1/n^2$,
this result is not modified for $8\pi^2\R(\S)>0.3$. For very small $\S$ 
we can make no statement because even higher orders of the power series
expansion become important.}
This means that indeed the only minimum is at $\S\to\infty$.

These models show that dynamical stabilization of the gauge coupling constant
is possible in supersymmetric vacua.
This mechanism determines $8\pi^2\vev\S$ at values of order one, but 
$\vev\S\gg(8\pi^2)^{-1}$ is only possible when some of the coefficients
of the polynomial $\cP$ (or  $\eta\,\xi^k$ for the model (\ref{W_DK}))
are exponentially small, which is not natural. Thus for moduli which
determine the gauge coupling of a non-perturbative
gauge group this mechanism is viable but
it is not so attractive for stabilizing
the dilaton of perturbative string theory.

Finally, let us observe that 
the results obtained for the $SU(2)$ gauge theory can easily be generalized
to other (symplectic or unitary) gauge groups. First consider 
an $Sp(2N_c)$ gauge theory with $N_f=N_c+1$ quark flavors, i.e.\ $2(N_c+1)$
quarks $Q_i$ in the fundamental representation of the gauge group. 
The low energy theory is described by a quantum moduli space spanned
by the VEVs of the `mesons' $M_{ij}=Q^r_iJ_{rs}Q^s_j$, where $J$ is the
$Sp(2N_c)$ invariant skew-symmetric form. They satisfy the constraint \cite{IP}
\be \Pf M\equiv {1\over2^{N_f}N_f!}\,\epsilon^{i_1\ldots i_{2N_f}}M_{i_1i_2}
                 \cdots M_{i_{2N_f-1}i_{2N_f}}=2^{N_c-1}\Lambda^{2(N_c+1)}.
\ee
The IYIT superpotential has the same form as in eq.\ (\ref{W_ITVY}),
\be W=\rho\,\Tr(MX)+U\ln{\Pf M\over\Lambda^{2(N_c+1)}},
\ee
where the factor $2^{N_c-1}$ has been absorbed in a redefinition of
the mesons.
The minimization of the supergravity potential is very analogous
to the $SU(2)$ case and yields
\be \label{V_Sp}
    \cV|_{\rm min}= (N_c+1)\,|\rho|^2\Mpl^4{\exp\left((N_c+1)\sum_{n=1}^\infty 
                  a_ne^{-8\pi^2\R(\S)n/(N_c+1)}
                  -{16\pi^2\over N_c+1}\R(\S)\right)\over 2\R(\S)^d}
    \ +\ \cO(|\rho|^4),
\ee
where we have used $\Lambda^{2(N_c+1)}=\Mpl^{2(N_c+1)}e^{-8\pi^2\S}$.

A generalization to $SU(N_c)$ gauge theories is technically slightly
more difficult because of the appearance of baryons, but this should not 
change the conclusions. The superpotential is (we need two additional 
singlets $Y$ and $\tilde Y$)
\be   W=\rho(\Tr(MX)+YB+\tilde Y\tilde B)
      +U\ln{\det M-\tilde BB\over\Lambda^{2N_c}}.
\ee
We expect that the supergravity potential will be qualitatively very
similar to (\ref{V_Sp}). Thus, the dilaton runaway problem is present
in all IYIT type models with unitary or symplectic gauge groups. 

Let us summarize.
In this paper we investigated the IYIT mechanism 
for the case of a dynamical, that is field dependent
gauge coupling. We argued that the presence of a
dilaton-like modulus $\S$ requires the coupling
of the IYIT model to supergravity.
This can be achieved by employing the
Veneziano--Yankielowicz formalism.
The minimization of the corresponding
supergravity potential shows that the vacuum
is destabilized and runs to the supersymmetric
configuration at $\Lambda =0$.
This result holds for any symplectic
(and presumably also unitary)
gauge theory with only fundamental matter.
Some more general tree-level superpotentials for the
mesons show non-trivial minima for $\S$ but these
are supersymmetric ground states. Thus, it seems that
for a gauge theory whose low energy dynamics is described
by a quantum modified moduli space only two situations generically occur
when a non-vanishing tree-level potential is present: 
Either in global supersymmetry
there is a stable ground state which leads to a supersymmetric
minimum for all fields including $\S$ in supergravity
or global supersymmetry is broken by the ground state but
restored when the gauge coupling gets dynamical because
$\vev\S$ is driven to infinity.

\vspace{1cm}

\begin{center} {\bf Acknowledgements} \end{center}
We would like to thank K.\ F\"orger,
V.\ Kaplunovsky and H.-P.\
Nilles for helpful discussions.
The work of M.K. is supported by the DFG. 
The work of J.L. is supported in part by GIF -- the German--Israeli
Foundation for Scientific Research
and NATO under grant CRG~931380.


\begin{thebibliography}{99}
\bibitem{PT} E.~Poppitz and S.~P.~Trivedi, \hep{9803107}.
\bibitem{IT} K.-I.~Izawa and T.~Yanagida, \Prog{95} (1996) 829, \hep{9602180};
   K.~Intriligator and S.~Thomas, \Nucl{473} (1996) 121, \hep{9603158}.
\bibitem{Seib_D49} N.~Seiberg,  \PhysR{49} (1994) 6857,
\hep{9402044}.
\bibitem{qmm1} C.~Cs\'aki, M.~Schmaltz and W.~Skiba, \PhysR{55} (1997) 7840,
   \hep{9612207}. 
\bibitem{qmm2} B.~Grinstein and D.R.~Nolte, \hep{9710001}.
\bibitem{kss} S.\ Kachru, N.\ Seiberg and E.\ Silverstein, \Nucl{480} (1996) 
   170,\\ \hep{9605036}.
\bibitem{ks} S.\ Kachru and E.\ Silverstein, 
\Nucl{482} (1996) 92,  \hep{9608194};\\
S.\ Kachru, 
%{\it ``Aspects of N=1 String Dynamics''},
\NuclProc{61} (1998) 42, \hep{9705173}.
\bibitem{mkL} M.~Klein and J.~Louis, \Nucl{511} (1998) 197, \hep{9707212}.
\bibitem{MBMQ} I.~Maksymyk, C.~P.~Burgess, A.~de~la~Macorra
and  F.~Quevedo, \hep{9712178}.
\bibitem{VY} G.~Veneziano and S.~Yankielowicz, \PhysL{113} (1982) 231;\\
   T.R.~Taylor, G.~Veneziano and S.~Yankielowicz, \Nucl{218} (1983) 493.
\bibitem{AKMRV} D.~Amati, K.~Konishi, Y.~Meurice, G.~C.~Rossi
and  G.~Veneziano, \PhysRep{162} (1988) 169.
\bibitem{DK} G.~Dvali and Z.~Kakushadze, \PhysL{417} (1998) 50, \hep{9709093}.
\bibitem{ADS} I.~Affleck, M.~Dine and N.~Seiberg, \Nucl{241} (1984) 493;
              \Nucl{256} (1985) 557.
%\bibitem{O_R} L.~O'Raifeartaigh, \Nucl{96} (1975) 331.
\bibitem{smallinst} E.~Witten, 
%{\it ``Small instantons in string theory''},
 \Nucl{460} (1996) 541, \hep{9511030}.
\bibitem{DS} M.~Dine and N.~Seiberg, \PhysL{162} (1985) 299.
\bibitem{W_index} E.~Witten, \Nucl{202} (1982) 253.
\bibitem{KL} V.~Kaplunovsky and J.~Louis, \Nucl{444} (1995) 191, \hep{9502077}.
\bibitem{int_in} K.~Intriligator, \PhysL{336} (1994) 409, \hep{9407106}.
%\bibitem{Seib_L318} N.~Seiberg, \hepp{9309335}, \PhysL{318} (1993) 469.
%\bibitem{Seib_B435} N.~Seiberg, \hep{9411149}, \Nucl{435} (1995) 129.
\bibitem{KS} A.~Kovner and M.~Shifman, \PhysR{56} (1997) 2396, \hep{9702174}.
\bibitem{CM} C.~Cs\'aki and H.~Murayama, \hep{9710105}.
\bibitem{SZ} M.~Schwetz and M.~Zabzine, \hep{9710125}.
\bibitem{BHOO} J.~de~Boer, K.~Hori, H.~Ooguri and Y.~Oz, \hep{9801060}.
\bibitem{IP} K.~Intriligator and P.~Pouliot, \PhysL{353} (1995) 471, 
   \hep{9505006}.

\end{thebibliography}
\end{document}